\documentclass[10pt]{article}

\usepackage{fancyhdr}
\usepackage{epsfig}
\usepackage{cite}
\usepackage{graphicx}
\usepackage[centertags]{amsmath}
\usepackage{theorem}

\usepackage{amssymb}
\usepackage{latexsym}
\usepackage{epic}
\usepackage{pstricks}
\usepackage{color}
\usepackage{mathrsfs}
\usepackage{latexsym}
\usepackage{epic}
\usepackage{braket}

\newcommand{\be}{\begin{equation}}
\newcommand{\ee}{\end{equation}}
\newcommand{\bea}{\begin{eqnarray}}
\newcommand{\nn}{\nonumber}
\newcommand{\eea}{\end{eqnarray}}
\newcommand{\ti}{\widetilde}
\usepackage{slashed}
\numberwithin{equation}{section}

\newcommand{\gl}{\lambda}
\newcommand{\refe}[1]{Eqn.~(\ref{#1})}

\renewcommand{\thefootnote}{\fnsymbol{footnote}}

\begin{document}
\center{
\begin{center}
{\Large\textbf{General Gauge Mediation in 5D}}
\vspace{1cm}

\textbf{Moritz McGarrie\footnote{m.mcgarrie@qmul.ac.uk} and  Rodolfo Russo\footnote{r.russo@qmul.ac.uk}}\\
\end{center}
\it{ Queen Mary University of London\\
Center for Research in String Theory\\
Department of Physics\\
Mile End Road, London, E1 4NS, UK.}}

\setcounter{footnote}{0}
\renewcommand{\thefootnote}{\arabic{footnote}}
\abstract{We use the ``General Gauge Mediation'' formalism to describe a 5D setup with an $S^{1}\!/\mathbb{Z}_{2}$ orbifold.  We first consider a model independent SUSY breaking hidden sector on one boundary and generic chiral matter on another.  Using the definition of GGM, the effects of the hidden sector are contained in a set of global symmetry current correlator functions and is mediated through the bulk.  We find the gaugino, sfermion and hyperscalar mass formulas for minimal and generalised messengers in different regimes of  a large, small and intermediate extra dimension. Then we use the 5D GGM formalism to construct a model in which an $SU(5)$ ISS model is located on the hidden boundary.  We weakly gauge a global symmetry of the ISS model and associate it with the bulk vector superfield. Compared to 4D GGM, there is a natural way to adjust the gaugino versus sfermion mass ratio by a factor $(M \ell)^{2}$, where $M$ is a characteristic mass scale of the SUSY breaking sector and $\ell$ is the length of the extra dimension.}

\section{Introduction}
Meade, Seiberg and Shih \cite{Meade:2008wd} constructed a $4D$ formalism to describe in general $\mathcal{N}=1$ gauge mediated supersymmetry breaking (GGM) in which, by construction, the hidden sector and MSSM completely decouple as the MSSM gauge couplings approach zero.  The formalism is valid for strongly coupled hidden sectors and all the supersymmetry breaking effects are encoded in a set of current correlator functions \cite{Buican:2008ws,deGouvea:1997tn,Intriligator:2008fr,Ooguri:2008ez,Distler:2008bt,Benini:2009mz,Buican:2009vv,Argurio:2009ge,Luo:2009kf,Kobayashi:2009rn,Lee:2010kb,Intriligator:2010be}. The decoupling of hidden and visible sector includes a large class of SUSY breaking models and the formulation highlights those properties that are generic to GGM and those that are model dependent. A particularly popular and natural form of gauge mediated supersymmetry breaking is the construction of ISS \cite{intriligator2006dsb}.  In ISS, supersymmetry is broken non-perturbatively in the electric description and is metastable.  It is a simple $\mathcal{N}=1$ SQCD model and as a result one may apply Seiberg duality to obtain an effective magnetic description in which supersymmetry breaking can be explored perturbatively.  It is well known that this theory has a signature of light gauginos and heavier sfermions  and this is seen as an unfortunate drawback.  In fact a recent paper on gauge mediation \cite{Abel:2009ve} (see also \cite{Rajaraman:2009ga}), scanned the parameter space of viable low energy MSSM spectra in terms of the scales  $\Lambda_{G}$ and $\Lambda_{S}$  of the high energy theory\footnote{These are the effective gaugino and effective scalar masses of the high energy theory.}.  They found that reasonable MSSM phenomenology could be obtained with an inverted hierarchy $\Lambda_{G}>\Lambda_{S}$  and this region was far larger than for $\Lambda_{G}<\Lambda_{S}$ and  $\Lambda_{G}\approx \Lambda_{S}$. Intriguingly, they suggest that there is no lower bound on the inital value of $\Lambda_{S}$. Their message is clear: Scalar terms like the $B_{\mu}$ term and scalar masses of the MSSM can be generated radiatively and that fermionic terms like gaugino masses are relatively insensitive to radiative corrections.  One may wonder, is there a framework in which one may naturally construct this inverted hierarchy and access the full parameter space of GGM?  The answer is gaugino mediation \cite{Mirabelli:1997aj,Kaplan:1999ac,Chacko:1999mi,Csaki:2001em}.

The gaugino mediated approach, which, in its crudest form\footnote{When messengers are coupled to a spurion multiplet $X=M+\theta^{2}F$, with $M^{2}\gg F$ , one may integrate out the messenger sector. The result is a pure gauge Lagrangian with holomorphic coupling in $\tau(X)$. Expanding in $\theta,\bar{\theta}$, one obtains the above result times a coefficient function of the gauge kinetic term~\cite{Giudice:1997ni}.}, one directly couples a spurion to the field strength of the vector superfield:
\begin{equation}
\int d^5 x \delta (x^5) \int d^{2} \theta  \frac{X}{M^{2}}  W^{\alpha}W_{\alpha}+h.c. \label{1point1}
\end{equation}
is still \emph{gauge mediated} and as a result of mediation in the bulk, this time the sfermions masses are suppressed and the gaugino masses dominate the soft terms.   Of course a realistic model of this type is likely to have a messenger sector (commonly fundamental messengers of $\mathcal{N}=1$ SQCD) and we expect to recover a bulk version of general gauge mediation. We would like to stress that in our calculations (which are at leading order in $g$) the gaugino and sfermion masses are generated by different current correlators and the sfermion mass is \emph{not} simply proportional to the square of the current correlator that generates the gaugino mass, as is implied by ``Gaugino Mediation'' \cite{Kaplan:1999ac,Chacko:1999mi}.  Applications of a 5D GGM, in the context of an AdS/CFT scenario have already been appreciated in the literature \cite{Benini:2009ff,McGuirk:2009am}.

In this paper we construct the 5D formalism for a general gauge mediation with an $S^{1}/\mathbb{Z}_{2}$ orbifold. In our analysis the susy breaking dynamics is confined to a 4d brane and thus the current correlators are exactly those of the usual GGM formalism and are independent of momentum in the 5th dimension.  
We apply the analysis of~\cite{Mirabelli:1997aj} in a GGM context and derive the scalar and fermion masses in terms of the current correlators. We briefly discuss Semi-Direct gauge mediation and then focus on the possibility of placing an ISS model on the hidden sector brane.

Section \ref{sec:Frame} outlines the decomposition of the $\mathcal{N}=2$ Non-Abelian bulk action, under orbifold conditions and is given in full in appendix \ref{NON}.  Section \ref{sec:3} looks at brane localised current correlators in the general gauge mediation framework and computes the general gaugino, sfermion and hypermultiplet susy breaking mass terms, the vacuum and Casimir energy and the semi-direct mass terms.  Section \ref{sec:general} produces explicit formulas for gaugino and sfermion susy breaking mass terms for a generalised hidden sector.  Section \ref{sec:brane} applies our results to an explicit hidden sector model by locating the ISS model on the hidden brane \cite{intriligator2006dsb}. In section \ref{sec:conclusion} contains our conclusions.  The Appendices include relevant calculations to obtain the main results in the paper and outlines our conventions.
\section{Framework}\label{sec:Frame}
In this section we recall the main features of $\mathcal{N}=1$ super Yang-Mills and hypermultiplet matter in $5d$. Once compactified on an orbifold of $S^{1}/\mathbb{Z}_{2}$, a positive parity vector multiplet couples to the boundaries of the orbifold and we associate this with the standard model gauge groups in the bulk.  The remaining fields fill a negative parity chiral multiplet which we do not couple to the boundaries. Similarly we will outline features of the orbifold compactified hypermultiplet.  A complete description is found~\cite{Hebecker:2001ke}, see also Appendix \ref{NON}. 

We first focus on the pure super Yang-Mills theory. The action written in components is
\be
S_{5D}^{SYM}= \int d^{5} x
~\text{Tr}\left[-\frac{1}{2}(F_{MN})^2-(D_{M}\Sigma)^2-i\bar{\gl}_{i}\gamma^M
D_{M}\gl^{i}+(X^a)^2+g_{5}\, \bar{\gl}_i[\Sigma,\gl^i]\right].
\ee
The coupling $1/g^2_{5}$ has been rescaled inside the covariant derivative, $D_{M}=
\partial_{M}+ig_{5} A_{M}$.  The other fields are a real scalar $\Sigma$, an $SU(2)_{R}$ triplet of
real auxiliary fields $X^{a}$, $a=1,2,3$ and a symplectic Majorana
spinor $\gl_{i}$ with $i=1,2$ which form an $SU(2)_R$ doublet. The reality condition is $\gl^i= \epsilon^{ij} C\bar{\gl}_{j}^{T} $.
Next, using an orbifold $S^{1}\!/\mathbb{Z}_{2}$ the boundaries will
preserve only half of the $\mathcal{N}=2$ symmetries. We choose to
preserve $\epsilon_{L}$ and set $\epsilon_{R}=0$.  We have a parity
operator $P$ of full action $\mathbb{P}$ and define
$P\psi_{L}=+\psi_{L}$ $P\psi_{R}=-\psi_{R}$ for all fermionic fields
and susy parameters. One can then group the
susy variations under the positive parity assignments and they fill an off-shell $4d$ vector
multiplet $V(x^{\mu},x_{5})$.  Similarly the susy variations of odd parity
form a chiral superfield $\Phi(x^{\mu},x_{5})$. We may therefore write a $5d$
$\mathcal{N}=1$ vector multiplet as a $4d$ vector and
chiral superfield:
\begin{alignat}{1}
V=&- \theta\sigma^{\mu}\bar{\theta}A_{\mu}+i\bar{\theta}^{2}\theta\gl-
i\theta^{2}\bar{\theta}\bar{\gl}+\frac{1}{2}\bar{\theta}^{2}\theta^{2}D\\
\Phi=& \frac{1}{\sqrt{2}}(\Sigma + i A_{5})+
\sqrt{2}\theta \chi + \theta^{2}F\,,
\label{fields}
\end{alignat}
where the identifications between $5d$ and $4d$ fields are
\begin{equation}
D=(X^{3}-D_{5}\Sigma) \quad F=(X^{1}+iX^{2})\,,
\end{equation}
and we used $\lambda$ and $\chi$ to indicate $\lambda_{L}$ and
$-i\sqrt{2}\lambda_R$ respectively. 

The bulk hypermultiplet action \bea
S_{5D}^{H} &=& \int d^5 x[-(D_MH)^\dagger_i(D^MH^i)-
i\bar{\psi}\gamma^MD_M\psi+ F^{\dagger i}F_i-g_5
\bar{\psi}\Sigma\psi+g_5 H^\dagger_i(\sigma^aX^a)^i_jH^j
\nonumber\\
&& +g_5^2 H^\dagger_i\Sigma^2H^i+ig_5\sqrt{2}\bar{\psi}
\lambda^i\epsilon_{ij} H^j-i\sqrt{2}g_5H^{\dagger}_{i}
\epsilon^{ij}\bar{\gl}_{j}\psi\,]\label{hyperaction2}
\eea 
decomposes into a positive and negative parity chiral superfield, $PH=+H$ and $PH^c=-H^c$:
\bea
H &=& H_1+\sqrt{2}\theta\psi_L+\theta^2(F_1+D_5H_2-g_5\Sigma H_2)\\
H^c &=& H^\dagger_2+\sqrt{2}\theta\psi_R+\theta^2(-F^{\dagger }_{2}-D_5
H^\dagger_1- g_5 H^\dagger_1\Sigma)\,.
\eea
With our conventions,the dimensions of ($H_{i},\psi,F_{i}$) are ($\frac{3}{2},2,\frac{5}{2}$).

The hypermultiplets are intriguing, as in the simplest case they only couple to the branes via the gauge coupling $g_5$ so they satisfy the framework of general gauge mediation. However, their soft masses are different from traditional brane localised matter and this might be relevant if they could play the role of the Higgs multiplet of the MSSM \cite{ArkaniHamed:2001mi,Hall:2001zb}.  In this paper we will simply compute the zero mode masses of a generic hypermultiplet.

In the next section we will locate a susy breaking hidden sector on one boundary of the orbifold.  We will encode the hidden sector into a set of current correlators, and use the positive parity vector multiplet to generate, a gaugino mass, construct loops across the bulk to generate sfermion masses on the other orbifold boundary and finally construct loops to generate a mass for the zero mode of the bulk hypermultiplets. 


\section{General Gauge Mediation for bulk and boundaries}\label{sec:3}
In this section we follow~\cite{Meade:2008wd} and use the formalism of
current correlators in a 5D orbifold $R^{1,3}\times S^1 \!
/\mathbb{Z}_{2}$ where supersymmetry is broken only on one of the two
planes at the end of the interval. The gaugino and sfermion mass are
written in terms of current correlators on the supersymmetry breaking
plane. Additionally we explore the hypermultiplet scalar and fermion
masses via the same set of current correlators.

In a supersymmetric gauge theory, global current superfields
$\mathcal{J^A}$ have the component form
\begin{equation}
\mathcal{J^A}=J^\mathcal{A}+i \theta j^\mathcal{A} -
i\bar{\theta}\bar{j}^\mathcal{A}-
\theta \sigma^{\mu} \bar{\theta} j_{\mu}^\mathcal{A} + 
\frac 1 2 \theta^2 \bar{\theta} \bar{\sigma}^{\mu} 
\partial_{\mu} j^\mathcal{A} - \frac 1 2 \bar{\theta}^2 \theta 
\sigma^{\mu} \partial_{\mu} \bar{j}^\mathcal{A} - 
\frac 1 4 \theta^2 \bar{\theta}^2 \square J^\mathcal{A}\;,
\end{equation}
which by definition satisfies the conditions
\begin{equation}
\bar{D}^2 \mathcal{J}^\mathcal{A}=D^2 \mathcal{J}^\mathcal{A}=0\;.
\end{equation}
This implies the usual current conservation on $j_{\mu}$:
$\partial^{\mu} j^\mathcal{A}_{\mu}=0$.  
We now gauge the global symmetry and couple the current to the vector
superfield with
\begin{equation}
\mathcal{S}_{int}=2g_{5}\!\int\! d^5x d^{4}\theta \mathcal{J} 
\mathcal{V}\delta(x_{5})= \!\int\! d^5x g_{5}(JD- \gl j \!-
 \!\bar{\gl} \bar{j}-j^{\mu}A_{\mu})\delta(x_{5})
\end{equation}
We may relate 4d brane localised currents as 5d currents by
$J_{5d}=J_{4d}\delta(x_{5})$.  The vector multiplet is five
dimensional but is written in 4 dimensions as $V(x^{\mu},x^{5})$ and
has been coupled to the boundary fields. The 5d coupling, $g_{5}$, has
mass dimension, $\text{Dim}[g_{5}]= \frac{(4-D)}{2}$.  In this
normalisation, following \refe{fullaction}, the mass dimensions of
$(A_{\mu},\Sigma,\gl_{i},X_{a})$ are $(3/2,3/2,2,5/2)$.  It follows
that 5d currents that couple to these fields,
$(J_{\mu},J_{\gl^{i}},J_{X^{a}})$ have mass dimension
$(4,7/2,3)$. $\delta(x_{5})$ carries a mass dimension $1$. We
explicitly insert the relation for the $D$ term and keep the auxiliary
fields $X^{3}$. The change of the effective Lagrangian to
$O(g_{5}^{2})$ is
\begin{alignat}{1}\label{E:ChangeL}
\delta \mathcal{L}_{eff}=-& g^{2}_{5}\tilde{C}_{1/2}(0) i \lambda \sigma^{\mu} \partial_{\mu} \bar{\lambda} - g^{2}_{5}\frac {1} {4} \tilde{C}_1(0) F_{\mu\nu} F^{\mu\nu}-g^{2}_{5}\frac {1}{ 2}(M \tilde{B}_{1/2}(0) \lambda \lambda + M \tilde{B}_{1/2}(0)\bar{\gl}\bar{\gl})\\ \nonumber & +\frac {1}{ 2 }g^{2}_{5}\tilde{C}_0 (0)(X^{3})^{2}+\frac {1}{ 2 }g^{2}_{5}\tilde{C}_0 (0)(D_{5}\Sigma)^{2}-g^{2}_{5}\tilde{C}_0 (0)(D_{5}\Sigma)X^{3} \\&+g_{5}^{2}\braket{J j^{\mu}}((D_{5}\Sigma)A_{\mu}-X^{3}A_{\mu})+\cdots\;\nonumber.
\end{alignat}
These are evaluated in the IR ($p^{\mu}_{\text{ext}}=0$).  When using these components to construct the diagrams in Figure 1, one must include the full momentum dependence. The $\tilde{B}$ and $\tilde{C}$ functions are related to momentum space current correlators, found below.  The last 4 terms require comment:  the first three of these replace the $D^{2}$ term, in the last line the current correlator is found to be zero \cite{Buican:2009vv}.  In position space, the current correlators can be expressed in terms of their mass dimension\footnote{Renormalised operators of conserved currents receive no rescalings $Z_{J}=1$ and no anomalous dimension $\gamma_{J}=0$} and some functions $C_{s}$ and $B_{\frac{1}{2}}$,
\begin{alignat}{1}
\braket{J(x,x_{5})J(0,x'_{5})}=&\frac{1}{x^{4}}C_0(x^{2}M^{2})\delta(x_{5})\delta(x'_{5}) \\
\braket{j_\alpha(x,x_{5})\bar j_{\dot\alpha}(0,x'_{5})}=&-i\sigma_{\alpha\dot\alpha}^\mu \partial_\mu(\frac{1}{x^{4}}C_{1/2}(x^{2}M^{2}))\delta(x_{5})\delta(x'_{5})\\
\braket{j_\mu(x,x_{5})j_\nu(0,x'_{5})}=&(\partial^2\eta_{\mu\nu}-\partial_\mu \partial_\nu)(\frac{1}{x^{4}}C_1(x^{2}M^{2}))\delta(x_{5})\delta(x'_{5})\\
\braket{j_\alpha(x,x_{5})j_\beta(0,x'_{5})}=&\epsilon_{\alpha\beta}\frac{1}{x^{5}}B_{1/2}(x^{2}M^{2})\delta(x_{5})\delta(x'_{5}) \\
\braket{j_\mu(x,x_{5})J(0,x'_{5})}=& cM^{2}\partial_{\mu}(\frac{1}{x^{2}}) \delta(x_{5})\delta(x'_{5}) \label{scale}
\end{alignat}
$M$ is a characteristic mass scale of the theory (e.g. the fermion mass of the SUSY breaking messenger multiplet).   $B_{1/2}$ is a complex function, $C_{s}$, $s=0,1/2, 1$, is real.
When supersymmetry is unbroken
\begin{equation}
C_0=C_{1/2}=C_1\;,\qquad \text{and} \qquad B_{1/2}=0\;.
\end{equation}
Supersymmetry is restored in the UV such that
\begin{equation} \label{BCUV}
\lim_{x \rightarrow 0} C_0(x^2 M^2)=\lim_{x \rightarrow 0} C_{1/2}(x^2 M^2) =\lim_{x \rightarrow 0} C_1(x^2 M^2)\;,\qquad \text{and} \qquad \lim_{x \rightarrow 0} B_{1/2}(x^2 M^2)=0\;.
\end{equation}
 $\tilde{C}_s$ and $\tilde B$ are Fourier transforms of $C_s$ and $B$,
\begin{equation}
\begin{split}
\tilde{C}_s\left(\frac{p^2}{M^2};\frac{M}{\Lambda}\right)&=\int d^4 x e^{ipx} \frac 1 {x^4} C_s(x^2 M^2)\\
M\tilde{B}_{1/2}\left(\frac{p^2}{M^2}\right)&=\int d^4 x e^{ipx} \frac 1 {x^5} B_{1/2}(x^2 M^2)\;.
\end{split}
\end{equation}
The $\tilde{C}_s$ and $\tilde B$ terms are the nonzero current correlator functions of the components of the current superfield.  The correlators  have positive parity ($P=+1$) as they live on the wall. Using the full action of $\mathbb{P}$, the Fourier transforms over $x_{5}$ and $x'_{5}$ removes the delta functions. In this off-shell formalism $\delta(0)$ \emph{does not} enter explicitly in the calculation (compare with \cite{Mirabelli:1997aj}). In momentum space we have,
\begin{alignat}{1}
\braket{J(p,p_{5})J(-p,p'_{5})} =&\tilde{C}_0(p^2/M^2)  \label{eq:c0}   \\
\braket{j_\alpha(p,p_{5})\bar j_{\dot\alpha}(-p,p'_{5})} =&-\sigma_{\alpha\dot\alpha}^\mu p_\mu\tilde{C}_{1/2}(p^2/M^2) \label{eq:c1/2}		\\ 
\braket{j_\mu(p,p_{5})j_\nu(-p,p'_{5})} =&-(p^2\eta_{\mu\nu}-p_\mu p_\nu)\tilde{C}_1(p^2/M^2)	\label{eq:c1}	\\ 
\braket{j_\alpha(p,p_{5})j_\beta(-p,p'_{5})} =&\epsilon_{\alpha\beta}M\tilde{B}_{1/2}(p^2/M^2)     \label{eq:b1/2}	\\ 
\braket{j_\mu(p,p_{5})J(-p,p'_{5})} =& c M^{2}\frac{2{\pi}^2 i p_{\mu}}{p^2}\label{eq:fsdf}
\end{alignat}
We see that the current correlators are completely independent of the momentum in the fifth dimension. The analysis of \cite{Buican:2009vv} demonstrates that $c=0$ in the last equation.

\subsection{Gaugino masses}
At $g^2$ order the susy breaking contribution to the gaugino mass can be read directly from the
Lagrangian \eqref{E:ChangeL} after rescaling $\lambda$ so as to
canonically normalise the bulk action~\eqref{fullaction}:
\begin{equation}\label{E:Gaugino}
M^{nm}_{\lambda } = g^2_{4} M \tilde{B}_{1/2} (0)\;.
\end{equation}
This result replaces the result \refe{1point1}, and in particular is correct even when $F$ is \emph{not} smaller than $M^2$ or when there are multiplet messenger scales (see appendix B.2 \cite{Dumitrescu:2010ha}).
These terms are of Majorana type and couple every Kaluza-Klein mode
with every other mode with the same coefficient. In addition we have
the usual Kaluza-Klein tower of masses ($p_{5}=\frac{n\pi}{\ell}$)
which are of Dirac type and mix $\lambda^L_n$ and $\lambda^R_n$.  The mass
eigenstates will be in general a linear combination involving
different Kaluza-Klein modes. This is similar in vein to the ``see-saw'' mechanism and for large $\ell$ the lowest mass eigenstate can become very light. This highlights that for bulk mediation, the scale $\Lambda_{G}$ is not a good scale and must be replaced by the lightest gaugino mass eigenvalue. We comment on three regimes: 
\begin{description}
\item[Small $\ell$] 
When the scale of the extra dimension $1/\ell$ is much bigger than the scales $\sqrt{F}$ and $M$ then we return to an effective 4d theory and the zero mode mass is given by~\eqref{E:Gaugino}.
\item[Intermediate $\ell$]
When $F\leq 1/\ell^2\ll M^2$ the susy breaking mass $M^{mn}_\lambda$ is of order $F/M$ and the K.K. mass is much bigger because $F/M \leq (1/M) (1/\ell^2) \ll 1/\ell$. In this case the gaugino mass is still given by~\eqref{E:Gaugino} and $\Lambda_{G}$ is a good scale.  
\item[Large $\ell$]
When  $1/\ell^2 \ll F$, then one must be careful and see how $F$ and $M$ scale.   For instance if $F \sim M/\ell$, which is possible in this regime, then there is a sizeble mixing between the various K.K. modes and the first mass eigenstate is lighter than $M^{nm}_\lambda$. If $M^{nm}_\lambda \gg 1/\ell$ the lightest gaugino eigenstate can have a \emph{much} lower mass than $M^{nm}_\lambda$ due to mixing with the tower of K.K. modes.  In this case $\Lambda_{G}$ is \emph{not} a good scale.
\end{description}

\subsection{Sfermion masses}
\begin{figure}[ht]
\centering
\includegraphics[scale=0.7]{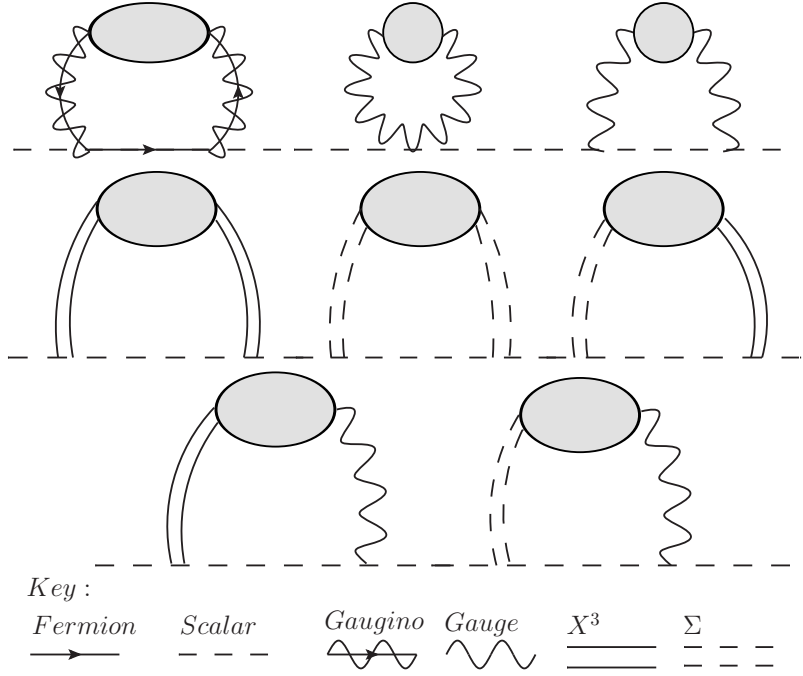}
\caption{The graphical description of the two point functions to the soft sfermion masses.  In the top row, the first diagram is from $\braket{j_{\alpha} \bar{j}_{\dot{\alpha}}}$ and the second and third are from $\braket{j_{\mu} j_{\nu}}$.  The middle row are all separately related to $\braket{J J}$, only the middle diagram survives propagation across the bulk. The final row is constructed from couplings to $\braket{J j_{\mu}}$ and are exactly equal to zero \cite{Buican:2009vv}.}
\label{od3}
\end{figure}

The sfermion masses can be determined in terms of the $\tilde{C}_{s}$ current correlator functions and propagation of the vector multiplet in the bulk.  This corresponds to the 8 diagrams in figure \ref{od3}.  The ``blobs'' are current correlators located on the hidden brane.  The scalar lines are located on the visible brane.  The intermediate propagators are the bulk fields and are components of the vector multiplet in the bulk.  The full momentum dependence of the current correlators must be taken into account as they form a part of a loop on the scalar propagator.  The first diagram is the dominant contribution in standard ``Gaugino Mediation''  \cite{Chacko:1999mi}.  The full set of diagrams are accounted for in \cite{Mirabelli:1997aj} including the two vanishing diagrams associated with $\braket{j_{\mu} J}$  (see also \cite{Buican:2009vv}). The top right most diagram contributes nothing to the mass due to transversality, when taking the external momentum to zero. The middle row has an auxiliary field $X^3$ which cannot propagate across the bulk so its diagrams vanish and only the middle one of that row survives.  In conclusion, when computing the soft mass terms, only the first two diagrams and the middle diagram of the middle row survive. They are the final ``supertraced'' combination with the same structure as in the 4D case.

To compute the three diagrams we need the propagator of a free massless bulk field 
\begin{equation}
  \left\langle{ a(x,x^5) a(y,y^5)} \right\rangle = 
	\int_{p5} \frac{i }{p^2 - (p_5)^2}\,
 e^{-ip\cdot (x-y)} (e^{ip_5(x^5-y^5)}
     + P e^{ip_5(x^5+y^5)}) \ ,
\label{fiveprop}\end{equation}
where 
\begin{equation}
   \int_{p5} = \int \frac{d^4p}{ (2\pi)^4} \frac{1}{ 2\ell}\sum_{p_5} \ ,
\label{intdef}\end{equation}
with $p_5$ summed over the values $ \pi n /\ell$, $n =$ integer. We propagate from $x^{5}=0$ to $y^{5}=\ell$.  The exponents of the 5th dimension, in brackets,  for two propagators will reduce to 
\begin{equation}
4(-1)^{n+\hat{n}}.
\end{equation}
In particular, this factor encodes the finite separation of the branes and will allow for a convergent finite answer for the soft mass.  It should also be noted neither brane (current correlator) conserves the incoming to outgoing $p_{5}$ momenta.  All vertex couplings can be determined by expanding out a canonical K\"ahler potential for a chiral superfield, which can be seen by example in Appendix \ref{generalised}.

We would like to factor out all the extra dimensional contributions to
the sfermions so that it leaves the GGM result multiplied by higher
dimensional contributions. We find
\begin{equation}
m_{\tilde{f}}^2= \sum _r g_{r(5d)}^4  c_2(f;r)E_r
\end{equation}
where
\begin{equation}
E_r= -\! \!\int\! \frac{d^4p}{ (2\pi)^4}\frac {1}{\ell^{2}}\sum_{n, \hat{n}} \!  \frac{(-1)^{n+\hat{n}}}{p^2-(p_{5})^{2}}\frac{p^{2}}{p^2-(\hat{p}_{5})^{2}} [3\tilde{C}_1^{(r)}(p^2/M^2)-4\tilde{C}_{1/2}^{(r)}(p^2/M^2)+\tilde{C}_{0}^{(r)}(p^2/M^2)], \label{Primeresult}
\end{equation}
where we used the regularisation of the K.K. sum described in Eq.(29) of~\cite{Mirabelli:1997aj}. $r=1,2,3.$ refer to the gauge groups $U(1),SU(2), SU(3)$. $c_2(f;r)$ is the quadratic Casimir for the representation of $f$ under the gauge group $r$. We have followed the convention of \cite{Intriligator:2010be} by using $E$, reserving $A$ for A-terms. The numerical coefficient in front of the $\tilde{C}_{s}$ terms in \refe{Primeresult} are essentially set by taking an index contraction of the current correlators \refe{eq:c0} to \refe{eq:c1}.  We use Matsubara frequency summation to identify
\begin{equation}
  \frac{1}{ \ell}\sum_{n}  (-1)^n \frac{1}{ k^2 + (k_5)^2}    
    = \oint \frac{dk^5}{ 2\pi} \frac{2 e^{ik_5 \ell}}{ e^{2ik^5\ell}-1} 
 \frac{1}{ k^2 + (k_5)^2}  = \frac{1}{ k}\frac{1}{ \sinh k\ell} .
\label{firstcontour}\end{equation} 
We obtain
\begin{equation}
E_r= \! -\!\int\! \frac{d^4p}{ (2\pi)^4}(\frac{1}{p\sinh p\ell})^{2} p^{2} [3\tilde{C}_1^{(r)}(p^2/M^2)-4\tilde{C}_{1/2}^{(r)}(p^2/M^2)+\tilde{C}_{0}^{(r)}(p^2/M^2)] 
\label{mainresult}
\end{equation}
All that is left is to evaluate the particular $E_{r}$ terms, which we will carry out for generalised messengers.  In the limit $\ell \to 0$, we have
\begin{equation}
\frac{1}{k}\frac{1}{ \sinh k\ell} \rightarrow \frac{1}{\ell k^{2}}
\end{equation}   
and we recover the 4d GGM answer \cite{Meade:2008wd}.

\subsection{Hypermultiplet scalar masses}
\begin{figure}[ht]

The supersymmetry breaking masses of the bulk hypermultiplet scalars and hypermultiplet fermions can also be computed in the gauge mediation setup and couple to the hidden brane exclusively via $g_5$, when using the action \refe{hyperaction2}. The diagrams for the scalars are similar to those of  Figure 1, but include an additional contribution with a bulk propagator, indicated with  the $\otimes$ symbol, coupling the positive parity gaugino to the negative parity bulk fermion 
\be
\braket{\gl_{L\alpha} \gl_{R\beta}}= \int \frac{d^4 k }{(2\pi)^4}\frac{1}{2\ell}\sum_{k_{5}}\frac{ i k_{5}\epsilon_{\alpha \beta}}{k^2-k^2_{5}}e^{-ik.(x-y)}(e^{ik_{5}.(x_{5}-y_{5})}+Pe^{-ik_{5}.(x_{5}+y_{5})}) 
\ee
A similiarly constructed propagator can be written in the $\sin (k_{5}x_{5})$ and $\cos (k_{5}x_{5})$ basis for the 5D wavefunction. 

\centering
\includegraphics[scale=0.8]{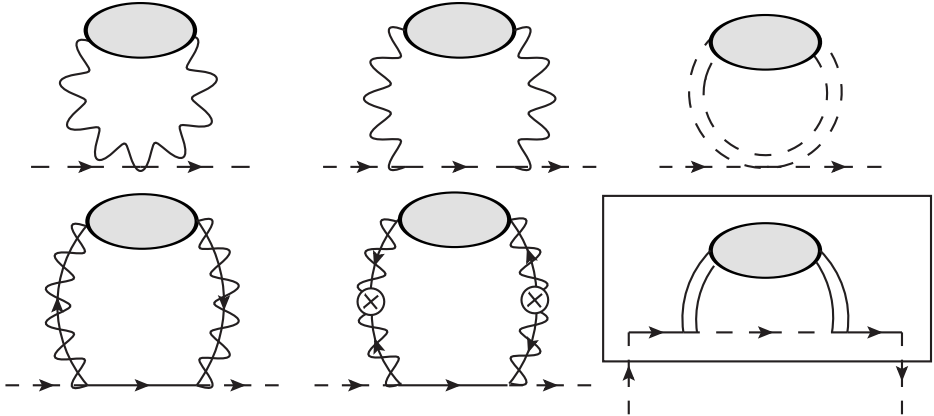}
\caption{The graphical description of the two point functions to the hypermultiplet scalar masses. Unlike the sfermion diagrams, the scalar field propagator does \emph{not} lie on a brane; the position of the vertex point must be integrated over for \emph{both} external sewing points when computing the diagrams. The parity of the external bulk scalar legs must also be specified.}
\end{figure}
To compute the diagrams: The current correlators (blobs) are brane localised, however the vertices joining the vector fields to the hyperscalar must be integrated over all of $y_{5}$. One must also specify the 5d wavefunction ($p_{5}$ momenta) of the external hyperscalar legs using
\be
\frac{1}{\sqrt{2\ell}}(e^{i\frac{n\pi}{\ell}y_{5}}+Pe^{-i\frac{n\pi}{\ell}y_{5}})\;\; (\text{for } n\neq 0), \;\;\; \;\frac{1}{\sqrt{\ell}} \;\; (\text{for } n=0)
\ee  
at the sewing point $y_{5}$, which is then integrated over. The second diagram does not contribute to the mass due to transversality. The rectangle in the final diagram represents that the diagram is completely localised on the hidden sector brane, including the vertices that couple to the external hyperscalar legs.

We will focus on the zero mode mass ($m_{0}\bar{H}_{0}H_{0}$)
\begin{equation}
m_{H_{0}}^2= \sum _r g_{r(5d)}^4 c_2(f;r)D_r
\end{equation}
where
\begin{equation}
D_r= -\! \!\int\! \frac{d^4p}{ (2\pi)^4}\frac {1}{2\ell^{2}}\sum_{n} \!  (\frac{p}{p^2+(p_{5})^{2}})^2 [3\tilde{C}_1^{(r)}(p^2/M^2)-4\tilde{C}_{1/2}^{(r)}(p^2/M^2)+\tilde{C}_{0}^{(r)}(p^2/M^2)].\label{secondresult2}
\end{equation}

\begin{equation}
D_r= \! -\!\int\! \frac{d^4p}{ (2\pi)^4}\frac{\coth (p\ell)+ p \ell \text{csch}^2(p \ell)  }{2 p \ell} [3\tilde{C}_1^{(r)}(p^2/M^2)-4\tilde{C}_{1/2}^{(r)}(p^2/M^2)+\tilde{C}_{0}^{(r)}(p^2/M^2)] 
\label{secondresult}
\end{equation}
The momentum integral is UV divergent. Physically this is to be expected as the hypermultiplet is not brane localised and so unlike the sfermion masses there is no brane separation to suppress large momenta contributions.  We can extract the $\ell$ dependent susy breaking mass  $\slashed{D}_{r}$ \cite{Buchbinder:2003qu}. We take $D_{r}= \slashed{D}_{r}+ \text{independent of $\ell$}$  where 
\be 
\slashed{D}_r= \! -\!\int\! \frac{d^4p}{ (2\pi)^4}\frac{\coth (p\ell)+ p \ell \text{csch}^2(p \ell) -1 }{2 p \ell} [3\tilde{C}_1^{(r)}(p^2/M^2)-4\tilde{C}_{1/2}^{(r)}(p^2/M^2)+\tilde{C}_{0}^{(r)}(p^2/M^2)]  \label{secondresult3}
\ee

Both in the case of bulk scalars and in the previous section on brane-localised scalars, we focused just on the case of external states at zero momentum. It would be interesting relax this condition and study, in a GGM setup, higher derivative operators by following the analysis of~\cite{Ghilencea:2005hm,Ghilencea:2006qm}.

\subsection{Vacuum Energy}
The propagation of supersymmetry in the bulk produces a non zero vacuum energy.   The computation of the vacuum diagrams at the order $O(g_5^2)$ is similar to the scalar masses in the previous sections. Similarly to the computation of the hypermultiplet scalar masses, the diagram with $X^3$ and that with $D_5 \Sigma$ combine and yield a contribution proportional to $p^2 \tilde{C}_0$. By including all other diagrams we obtain
\begin{equation}
E_{Vac}/V_{4d}= \frac{1}{4} g^{2}_{5}  d_{G} \int \frac{d^{4}p}{(2\pi)^{4}} \frac{p}{\tanh (p\ell)}[3\tilde{C}_1^{(r)}(p^2/M^2)-4\tilde{C}_{1/2}^{(r)}(p^2/M^2)+\tilde{C}_0^{(r)}(p^2/M^2)]. 
\end{equation}
$d_{G}$ is the dimension of the adjoint representation of the gauge group $r$. The vacuum energy is also UV divergent. The Casimir energy is the component of the vacuum energy where the bulk propagation winds $x^{5}$. This, in general, will contribute to the determination of the physical value of $\ell$, along with other supergravity corrections.  The Casimir energy is given by
\begin{equation}
\frac{E_{\text{Cas}}}{V_{4}}= \frac{1}{2}g^{2}_{5} d_{G}\int \frac{d^{4}p}{(2\pi)^{4}}\frac{p}{ e^{2p \ell}-1} [3\tilde{C}_1^{(r)}(p^2/M^2)-4\tilde{C}_{1/2}^{(r)}(p^2/M^2)+\tilde{C}_0^{(r)}(p^2/M^2)].\label{casimirenergy}
\end{equation}
\subsection{Semi-Direct Gauge Mediation via the bulk}
In \cite{Argurio:2009ge} semi-direct gauge mediation in a $4d$ setup is explored using current correlators. In this section we comment on the semi-direct case in our $5d$ setup with two $4d$ Branes and a $5d$ bulk.  One brane is the MSSM brane, described by some generic chiral matter, charged under the visible gauge group $G_{v}$, which lives in the $5d$ bulk.  The other brane is a SUSY breaking brane.  The messenger fields are located on this brane and are charged under both $G_{v}$ and a brane localised gauge group $G_{h}$.  The messengers do not participate directly in the susy breaking dynamics, however they couple to the brane localised susy breaking sector via gauge interactions with gauge group $G_{h}$ and by construction the messengers and susy breaking sector decouple as $g_{h}\rightarrow 0$.   

The Gaugino masses vanish at leading order (three loops) precisely because of the argument of \cite{Argurio:2009ge}. 
The sfermion masses in a flat bulk are found to be
\begin{equation}
m_{\tilde{f}}^2= \sum _r g_{r(5d)}^{(v)4} g_{r(4d)}^{(h)4} c_2(f;r)E_r
\label{useme}
\end{equation}
where
\begin{equation}
E_r= +\! \!\int\! \frac{d^4p}{ (2\pi)^8}\frac {1}{\ell^{2}}\sum_{n, \hat{n}} \!  \frac{(-1)^{n+\hat{n}}}{p^2-(p_{5})^{2}}\frac{p^{2}K(p^{2}/m^{2})}{p^2-(\hat{p}_{5})^{2}} [3\tilde{C}_1^{(r)}(p^2/M^2)-4\tilde{C}_{1/2}^{(r)}(p^2/M^2)+\tilde{C}_{0}^{(r)}(p^2/M^2)].
\end{equation}
 $m$ is the mass of the messengers. $K_{s}(p^{2}/m^{2})$ are the kernels, which in principle could be different for each of $s=0,1/2,1$. In \cite{Argurio:2009ge}, it was checked that $K_{0}=K_{1/2}=K_{1}$, so we will ignore this supscript index.  
 
As a final comment, one motivation for GGM5D is that it makes the partitioning of the hidden and visible sector a geometric feature.  One may be motivated to make semi-direct mediation a geometric feature too by placing the SUSY breaking sector $X$, the messengers $\phi,\phi^\dagger$ and the MSSM on three distinct branes. It would be interesting to study explicitly if it is possible to realise such a possibility in a concrete model.  
\section{Generalised Messenger sector}\label{sec:general}
In this section we give a concrete description of the
4-dimensional susy breaking brane and consider two sets of chiral
messenger $\phi_i,\ti\phi_{i}$ coupled to a spurion field $X$. We
follow~\cite{Marques:2009yu} and extend the usual setup of a
generalised messenger sector to the case where the gauge multiplet
propagates in a 5d orbifold.

The superpotential describing the coupling of the messengers and the
spurion is identical to that considered in~\cite{Marques:2009yu} and
is localised in the fifth dimension on the susy breaking brane
\be W_{\phi} = {\cal M}(X)_{ij}\ \phi_i \ti \phi_j = (m + X
\lambda)_{ij}\ \phi_i \ti \phi_j \label{superpotential}\ee
where $m$ and $\gl$ are generic matrices.  We assume that all chiral
fields have canonical kinetic term and so, after a field redefinition,
we can take ${\cal M}=m+\langle X \rangle \lambda$ to be diagonal with
real eigenvalues $m_{0k}$, where, as usual, $\langle X \rangle$ is the
vev of the scalar component of the spurion superfield $X =\langle X
\rangle+  \theta^2 F  $. Further, we take $F\lambda$ to be hermitian and
by using unitary matrices one may diagonalise the bosonic mass-squared
matrix
\be
\mathcal{M}^{2}_{\pm}=U^{\dagger}_{\pm}(\mathcal{ M}^{2} \pm
F\lambda)U_{\pm} \ee Such that $\mathcal{M}^{2}_{\pm}$ has real
eigenvalues $m^{2}_{\pm k}$.  We define two mixing matrices: \be
A^{\pm}_{kn}=(U^{\dagger}_{\pm} )_{kn}(U_{\pm} )_{nk} \quad \quad
B^{\pm}_{kn}=(U^{\dagger}_{\pm}U_{\mp})_{kn}
(U^{\dagger}_{\mp}U_{\pm} )_{nk}
\ee
The calculations are carried out explicitly in section \ref{generalised}; in this section we simply display the results.
\subsection{Gaugino masses}

As we have seen in the previous section, the Majorana gaugino mass
matrix of the 5d model couples every Kaluza-Klein mode to every other
with the same coefficient and this contribution is
captured in~\eqref{E:Gaugino}.
In this case we can compute explicitly the correlator determining
$M \ti {B}_{1/2}$ by using~\eqref{supercurrent}

\be M_{r} = g^2 M \tilde{B}_{1/2}(0) =\frac{\alpha_r}{4 \pi}
\Lambda_{G} \ , \ \ \ \ \Lambda_{G} = 2 \sum_{k,n = 1}^N \sum_\pm \pm\ d_{kn}\ A_{kn}^\pm\ m^0_n 
 \frac{(m_k^\pm)^2 \log ((m_k^\pm)^2/(m_n^0)^2)}{(m_k^\pm)^2 - (m_n^0)^2}. \label{44}\ee
$k,n$ are messenger indices running from $1$ to $N$, the number of messengers, while $d_{kn}$ is nonzero and equal to $d_{k}$ or $d_{n}$ only if $\phi_{n}$ and $\tilde{\phi}_{k}$ are in the same representation.  In the full mass matrix one must take into account the Dirac masses of
the Kaluza-Klein tower itself. However, the susy breaking
contribution~\refe{44} is identical to the purely 4-dimensional case
and it is possible to follow~\cite{Marques:2009yu} for various case by
case simplifications. For instance, when $F\ll M^2$, to lowest order in $F/M^{2}$ and for
$SU(N)$ fundamentals one finds
\be
\Lambda_{G}=\sum^{N}_{k=1}\frac{F\gl_{kk}}{m^0_{k}}=F \partial_{X}\log \text{det}
\mathcal{M}
\ee
which is a familiar $4d$ result.
\subsection{Sfermion masses}\label{sfermionmasses}
The sfermion masses are sensitive to the extra dimension $\ell$.  In the small $\ell$ limit the 4d results are recovered and this is fully explored in \cite{Marques:2009yu}. 
\subsubsection{Large $\ell$}
When $1/\ell^2$ is smaller than the scales $F$ and $X^{2}$  the sfermions masses can be written as 
\be m_{\ti f}^2 = 2 \sum_{r =1}^3 C_{\ti f}^r  \ \left(\frac{\alpha_r}{4\pi}\right)^2 \ \Lambda_{S}^2 
\label{alpha}
\ee
$C_{\ti f}^r$ are the quadratic Casimirs of $\ti f$ in the gauge group $r$. The sfermion scale $\Lambda_S^2$ is\footnote{See Appendix \ref{generalised} for its derivation.}
\bea 
\Lambda_S^2  =\sum_{k,n} \frac{\zeta (3)}{\ell^2}  \sum_{\pm}d_{kn}\!\!\! &&\!\!\![B^{\pm}_{kn}[\frac{2m^{2}_{\pm k}}{m^{2}_{\pm k}-m^{2}_{\mp n}}\log m^2_{\pm k}-1] + \delta_{kn}[\log m^{2}_{\pm k}m^4_{0k}] \nn\\
\!\!\! &&\!\!\!-\frac{4A^{\pm}_{kn}}{(m^{2}_{\pm k}-m^{2}_{0 n})}[m^2_{\pm k}\log m^2_{\pm k}-m^2_{0 n}\log m^2_{0 n}-1]     \label{LambdaS2}\\
\!\!\! &&\!\!\!-\frac{2A^{\pm}_{kn}}{(m^{2}_{\pm k}-m^{2}_{0 n})^2}[ (m^2_{\pm k}-m^2_{0 n})(m^2_{\pm k}+m^2_{0 n})-2 m^2_{\pm k}m^{2}_{0 n}\log \frac{m^2_{\pm k}}{m^2_{0 n}}]] \nn \eea
We may reduce to minimal gauge mediation \cite{Martin:1996zb} by setting $m = 0$ in equation (\ref{superpotential})
\be 
\Lambda_S^2 =  \left(\frac{F}{X}\right)^2 \left(\frac{1}{X \ell}\right)^2\sum_{k=1}^N    \zeta(3)
d_{kk} h(x_k)\label{LSMGM}\ee
\be
x_{k}=\frac{F}{\lambda_{k}X^2}
\ee
\be h(x) =\frac{3}{2}[\frac{4+x-2x^2}{x^4}\log(1+x)+\frac{1}{x^2}] + (x\rightarrow -x)
\ee
$h(x)$ for $x<0.8$ can be reasonably approximated by $h(x)=1$ and $\gl_{k}$ are the eigenvalues of $\gl$ in \eqref{superpotential}.
\begin{figure}
\begin{center}
\includegraphics[scale=1]{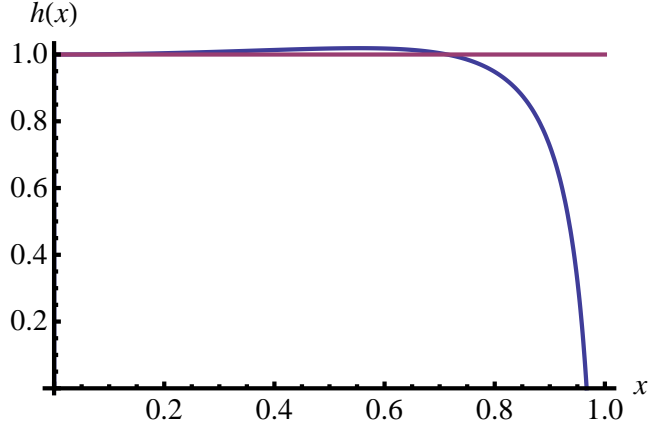}
\end{center}
\caption{A plot of the function $h(x)$ between  $x=0$ and $x=1$.}
\end{figure}
\newline
The limit of small multi-messenger mixing effects  gives 
\be \Lambda_S^2 =   \sum_{k = 1}^N  \zeta(3) d_{kk}\ \frac{F^2 \lambda_{k}^2}{\ell^2(m_k^0)^4} \ h\left(\frac{F \lambda_{k}}{(m_k^0)^2}\right) \label{ExpansionFSfermions}\ee
In the $h(x)=1$ limit, and small multi messenger mixing, we would like to derive the analog sfermion formula found in \cite{Cheung:2007es}.  The additional $|\ell^{2} M_{i}|^{2}$ factors cannot be taken inside such that we find
\be
\Lambda_{S}^{2} =\frac{\zeta(3)}{2}\sum_{i=1}^{N}\frac{|F|^{2}}{|\ell \mathcal{M}_{i}|^{2}}\frac{\partial^{2}}{\partial X \partial X^{*}}(\ln |\mathcal{M}_{i}|^{2})^{2}
\ee
Where $\mathcal{M}_{i}$ are in this case the complex eigenvalues of $\mathcal{M}$.  
\subsubsection{Intermediate $\ell$}
In the intermediate limit that $F\leq 1/\ell^2 \ll M^2$, the large $\ell$ results are still valid and $h(x)=1$. 
Reducing to minimal gauge mediation ($W=X\Phi\bar{\Phi}$) we find
 \be
\Lambda_S^2 = \frac{1}{\ell^2} 
\frac{2 F^2}{m^4}   \zeta(3)\sum_{k=1}^{N}d_{kk}
\ee
The limit of small multi-messenger mixing effects gives
 \be
\Lambda_S^2 = \frac{1}{\ell^2}\sum_{k=1}^{N}d_{kk} 
\frac{2 F^2\gl^{2}_{k}}{ (m^{0}_{k})^4} \zeta(3)
\ee

Finally we comment on the ratio $\frac{\Lambda^2_{G}}{\Lambda^{2}_{S}}$. In ``(Extra) Ordinary Gauge Mediation'' \cite{Cheung:2007es}, this quantity is defined as $N_{eff}$ and it may vary continously between $0$ and $N$, the number of messengers.  This definition is peculiar to the 4d models; we may easily  have $\Lambda_{S}^{2}\rightarrow 0$ and $N_{eff}\rightarrow \infty$ in this 5D construction.  To avoid confusion we will refer to it as a ratio and not as $N_{eff}$.

\subsection{Hyperscalar zero mode mass}
The positive parity scalar of the hypermultiplet have a susy breaking mass term. In this paper we compute only the zero mode scalar.  We may use the same expansions of the function in square brackets (Appendix \ref{sec:cterms}) we used for the sfermion masses.  In general we can define 
\be m_{H_{0}}^2 = 2 \sum_{r =1}^3 C_{H}^r  \ \left(\frac{\alpha_r}{4\pi}\right)^2  \Lambda^2_{H_{0}}.
\label{beta}
\ee
$C_{H}^r $ is the Casimir of the representation $r$ of the positive parity hypermultiplet $H$.  We now look at the various limits. 
\subsubsection{Small $\ell$}
In the small $\ell$ limit we start with \refe{secondresult2} and truncating the tower we find 
\begin{equation}
D_r= -\! \!\int\! \frac{d^4p}{ (2\pi)^4}\frac{1}{p^2} [3\tilde{C}_1^{(r)}(p^2/M^2)-4\tilde{C}_{1/2}^{(r)}(p^2/M^2)+\tilde{C}_{0}^{(r)}(p^2/M^2)].\label{secondresult4}
\end{equation}
This  is exactly the 4d result for sfermion masses found in \cite{Meade:2008wd}.  We may use all the generalised messenger results of \cite{Marques:2009yu}, with the identification
\be
\Lambda^{2}_{H_{0}} = \Lambda^2_{S (4d)}
\ee
where $\Lambda^2_{S (4d)}$ is the $4d$ result found in \cite{Marques:2009yu}.
\subsubsection{Large $\ell$}
When $1/\ell^2$ is smaller than the scales $F$ and $X^{2}$ we may start from \refe{secondresult3}. We know that the function in square brackets is independent of $p$ and can be found in appendix \ref{sec:cterms},  so we may evaluate the $p$ dependent integral independently. The result is 
\be
\Lambda^2_{H_{0}} =\frac{2}{3}\Lambda^2_{S}
\ee
$\Lambda^2_{S}$ is written explicitly in the previous subsection. The intermediate $\ell$ limit can be found by setting $h(x)=1$.  The zero mode fermions of the hypermultiplet receive no susy breaking mass corrections at order $g^4_{5d}$.
\subsection{Casmir Energy}
The Casimir energy can be computed for the generalised messenger sector just as for the sfermion and hyperscalar zero mode mass by using the results of appendix \ref{generalised}. In particular using \refe{casimirenergy} and restricting to the case of minimal gauge mediation and for a single set of fundamental messengers we find
\be
 E_{\text{Cas}}/V_{4d}=- g^2_{4} \frac{d d_{g} \zeta(5)}{512\pi^4} \frac{F^2_{X}}{(X \ell)^4}h(x).\label{vacua}
\ee
This result reproduces exactly the result of \cite{Mirabelli:1997aj}.  
\section{ISS on the brane}\label{sec:brane}
In this section we turn to an explicit application of 5D general gauge mediation.  We construct a scenario in which the ISS model   \cite{intriligator2006dsb,Intriligator:2007py} is located on the hidden brane.  We choose to explore supersymmetry perturbatively in the macroscopic (magnetic) variables.  We have an $\mathcal{N}=1$ SQCD with magnetic gauge group $SU(N)$  and $N_{f}$ flavours.  We weakly gauge a global symmetry and associate this with the bulk gauge field. The superpotential is
\begin{equation}
W_{\text{ISS}}= h\text{Tr}\tilde{\varphi}\Phi\varphi-h \text{Tr}[\mu^{2}\Phi]+ [\text{Deformations}]
\end{equation}
The magnetic meson $\Phi$ is a gauge singlet and an adjoint of the flavour group\footnote{This $\Phi$ should not be confused with the bulk $\Phi$ used in \refe{fields}.}. The magnetic quarks $\varphi$ and $\tilde{\varphi}$ are fundamental (antifundamental) of the gauge group and antifundamental (fundamental) of the flavour group, respectively.  Supersymmetry is broken by rank condition when $N_{f}>N$, as only the first $N$  F terms of $\Phi$ can be set to zero. Following the model explored in \cite{Kitano:2006xg,Zur:2008zg,Anguelova:2007at,Koschade:2009qu}, the matrix $\mu$ is explicitly broken
\begin{equation}
\mu^2_{AB} = \left(
\begin{array}{cc}
m^2 \mathbb{I}_{N} & 0  \\
0 & \mu^2 \mathbb{I}_{N_F-N} 
\end{array} \right)_{A B}
\label{t}
\end{equation}
with $\mu<m$, and  additionally we include the deformation
\be 
\delta W= h^2 m_{z} \text{Tr} \tilde{Z}Z.
\ee 
As mentioned mentioned more fully in \cite{Kitano:2006xg,Zur:2008zg,Anguelova:2007at,Koschade:2009qu}, this deformation explicitly breaks R-symmetry thus allowing gaugino masses.   The final unbroken vacuum symmetry groups and matter content is 
\begin{center}
\begin{tabular}{|ccc|}
\hline
Field & $SU(N)_{D} $& $SU(N_{f}-N)_{f} $\\
\hline
$\Phi = \left(
\begin{array}{cc}
Y_{\text{{\tiny $N$x$N$}}} & Z_{\text{{\tiny $N$x($N_{f}$-$N$)}}}  \\
\tilde{Z}_{\text{{\tiny ($N_{f}$-$N$)x$N$}}} & X_{\text{{\tiny ($N_{f}$-$N$)x($N_{f}$-$N$)}}}
\end{array} \right)_{\text{{\tiny $N_{f}$x$N_{f}$}}} $ &

$\left(
\begin{array}{cc}
\text{$\mathbf{Adj}+1$} & \bar{\square} \\
\square & 1
\end{array} \right) $ &

$\left(
\begin{array}{cc}
1 & \square \\
\bar{\square} & \text{$\mathbf{Adj}+1$}
\end{array} \right)$\\

$\varphi=\left ( \begin{array}{c}
\chi_{N \times N} \\
\rho_{N_{f}-N\times N}
\end{array} \right)_{\text{\tiny{NfxN}}}$

&$\left (\begin{array}{c}
\text{$\mathbf{Adj}+1$}\\ 
\square \end{array} \right)$ 

& $\left(\begin{array}{c}
1 \\
 \bar{\square}
\end{array} \right)$\\

$\tilde{\varphi}=\left ( \begin{array}{c}
\tilde{\chi}_{N \times N} \\
\tilde{\rho}_{N\times N_{f}-N }
\end{array} \right)_{\text{\tiny{NxNf}}}$

&$\left (\begin{array}{c}
\text{$\mathbf{Adj}+1$}\\ 
\bar{\square} \end{array} \right)$ 

& $\left(\begin{array}{c}
1 \\
 \square
\end{array} \right)$ \\
\hline 
\end{tabular}
\end{center} 
The superpotential is
\be
W= h\text{Tr}(\tilde{\chi} Y \chi + \tilde{\rho}Z\chi + \tilde{\chi}\tilde{Z}\rho+\tilde{\rho}X \rho)-h^2 m^2 \text{Tr}Y -h^2 \mu^2 \text{Tr}X+h^2m_{z}\text{Tr}\tilde{Z}Z.\label{super2}
\ee
Next, we choose to weakly gauge either of the flavour symmetry groups $SU(N)_{f}$ or $SU(N_{f}-N)_{f}$ and associate it with the gauge group in the bulk. For instance one may choose an $SU(5)$ standard model ``parent'' gauge group. We take these fields to have a canonical K\"ahler potential and all the matter on the MSSM brane to have a canonical K\"ahler potential coupled to this bulk gauge superfield. Classically the potential is 
\be
V_{ISS}=(N_{f}-N)|h^2\mu^4| .
\ee
The field $X$ is a classical modulus and its vacuum expectation value, $X_{0}$ is found by minimising
\begin{equation}
V_{Total}= V_{ISS}+ V_{CW}
\end{equation}
where $V_{CW}$ is the corresponding Coleman-Weinberg potential
\begin{equation}
V_{CW} = \frac{1}{64\pi^{2}}\text{STr}\mathbf{M}^4\text{Log}\frac{\mathbf{M}^{2}}{\Lambda^{2}}=
\frac{1}{64\pi^{2}}(\text{Tr}\mathbf{m}^{4}_{B}\text{Log}\frac{\mathbf{m}_{B}^{2}}{\Lambda^{2}}-
\text{Tr}\mathbf{m}^{4}_{F}\text{Log}\frac{\mathbf{m}_{F}^{2}}{\Lambda^{2}}).
\end{equation}
We find
\begin{equation}
X_{0}=\braket{X} = \frac{1}{2} h m_z, \; \; \; \; M^{2}_X = \frac{h^{4}\hat{\mu}^{2}}{12\mu^{2}\pi^{2}}\left(
\begin{array}{cc}
\hat{\mu}^{2}&  -\frac{9}{40}{X_0}^{2}  \\
-\frac{9}{40}{X_0}^{2} &\hat{\mu}^{2}
\end{array} \right)
\end{equation}
where we have expanded to first order in $h, m_z$ and in $\hat{\mu} / \mu$ up to first non-vanishing order \cite{Koschade:2009qu}. We have supressed factors of $N(N_{f}-N)$
in the expression for $M^{2}_X $ coming from tracing over degenerate mass eigenavalues.

\subsection{Large $\ell$ mediation} 
The susy breaking contribution to the gaugino masses are
\be
m^{n \hat{n}}_{\gl} =\frac{\alpha}{4\pi}\Lambda_{G}=\frac{\alpha}{4\pi}F_{X}\sum_{i}\frac{\partial_{X}M_{i}}{M_{i}} \label{nice}
\ee
where $M_{i}$ are the eigenvalues of the fermion mass matrix of messengers derivable from \refe{super2} and computed in \cite{Zur:2008zg,Koschade:2009qu}.  The messenger sector is dominantly $(\rho,Z)$ in both embeddings of the standard model and we find the susy breaking mass to be
\be 
\Lambda_{G}= \frac{N h^2 \mu^2 m_{z}}{(m^2-hX_{0}m_{z})}.
\ee 
As we have highlighted throughout, the mass eigenstates must be found after inclusion of the K.K. masses, to the mass matrix.  The sfermion masses can be found, using the results of the previous section, from
\be
m^{2}_{\tilde{f}} =2 C_{\tilde{f}}(\frac{\alpha}{4\pi})^{2}\Lambda^{2}_{S}
\ee
with
\be
\Lambda_{S}^{2}= \frac{N \zeta(3)}{\ell^2} |F_{X}|^{2} \sum_{i}|\frac{\partial_{X}M_{i}}{M^{2}_{i}}|^{2}\label{moremasses}
\ee
We find
\be
\Lambda_{s}^{2}= \frac{\zeta(3)N \mu^4}{\ell^{2}}\frac{[2m^6 \!+\! 2h^3 m^2 m_{z}^{3}(3hm_{z}-2X)\!+\! h^4 m^{4}_{z}(-hm_{z}+X_{0})^{2}+m^4(9h^2m^2_{z}+2hm_{z} X_{0}\!+\! X^2_{0})]}{(m^2-hm_{z}X_{0})^4(4m^2+(X_{0}-hm_{z})^2)}
\ee
Which is non-zero but highly suppressed. 

\subsection{Intermediate $\ell$ mediation} 
In the intermediate range that $F\leq 1/\ell^2\ll M^2$ the vacuum energy is given by \refe{vacua}, with $h(x)=1$ This time, however, the zero mode gaugino mass \emph{can} be approximated by \refe{nice}: 
\be 
m^{0}_{\gl}= \frac{N \alpha h^2 \mu^2 m_{z}}{4\pi(m^2-hX_{0}m_{z})}.
\ee
and there is no problem of K.K. mixing.  The sfermion masses are still given by \refe{moremasses}. As a first approximation, one may ignore this contribution to sfermion masses entirely.  One then lets the renormalisation group flow at the high scale down to the low scale generate sufficiently heavy sfermions to avoid current bounds.

\section{Conclusion and Discussion}\label{sec:conclusion}
We have computed bulk gauge mediated supersymmetry breaking of $\mathcal{N}=1$ supersymmetry from a hidden to a visible brane, using the current correlator techniques of \cite{Meade:2008wd}. We obtained analytic results for both semi-direct and direct coupling of the messengers and susy breaking sector, located on a hidden brane. 
We then used the ISS model \cite{intriligator2006dsb} as an explicit example of a brane localised hidden sector that can be accurately described by this formalism.
In particular we would like to highlight that our formalism was off-shell and naturally included otherwise difficult $\delta(x_{5})$ terms.  Typically a small extra dimension $\ell$ returns this model to the effective $4d$ ``General Gauge Mediation''.  So for ``Gaugino Mediation'' to play a role $\ell$ should be large compared to the mass scale of the susy breaking theory.  We found that the susy breaking gaugino masses mix with the Kaluza-Klein masses when the two mass scales are comparable (i.e. for $1/\ell^2$ smaller than $F,M^2$), resulting in a light gaugino zero mode.  To circumvent this one must choose the susy breaking Majorana mass to be at least of order $10$ times larger than the $n=1$ K.K. mass.  We therefore exploit and highlight an intermediate regime in which $F\leq 1/\ell^2 \ll M^2$. In this case the zero mode gaugino is mostly the susy breaking Majorana contribution and the sfermions are still suppressed. 

A preliminary analysis of a latticised construction of general gauge mediation following the lines of \cite{Csaki:2001em,ArkaniHamed:2001ca}, has been carried out and it replicates the characteristic features of the GGM5D soft terms found in this paper. This also confirms results found in \cite{Csaki:2001em}.  

Finally, we would like to point out that a full scan of the $5D$  GGM parameter space is also worthwhile, especially as we have shown that $\Lambda_{G}$ is not a good scale in all parts of the parameter space and so does not coincide with the $4D$ scans \cite{Abel:2009ve,Rajaraman:2009ga}.

\paragraph{Acknowledgments} 
We would like to thank Steven Thomas, Daniel Koschade, Daniel
C. Thompson, Diego Marqu\'es and Mathew J. Dolan for useful
discussions. This work is partially supported by STFC under the
Rolling Grant ST/G000565/1.  MM is funded by STFC.

\appendix

\section{Non-Abelian Bulk Action}\label{NON}
This appendix reviews the $\mathcal{N}=1$ 5D Non-Abelian bulk action.  This corresponds to $\mathcal{N}=2$ in the 4D perspective.  We compactify on an orbifold, $S^1/\mathbb{Z}_{2}$, such the pure super-Yang-Mills becomes a  $\mathcal{N}=1$  positive parity vector multiplet and negative parity chiral multiplet.  This review is closely based on ~\cite{Hebecker:2001ke,Mirabelli:1997aj}.  

Starting with the $\mathcal{N}=1$ pure super-Yang-Mills in components
\be
S_{5D}^{SYM}= \int d^{5} x
~\text{Tr}\left[-\frac{1}{2}(F_{MN})^2-(D_{M}\Sigma)^2-i\bar{\gl}_{i}\gamma^M
D_{M}\gl^{i}+(X^a)^2+g_{5}\, \bar{\gl}_i[\Sigma,\gl^i]\right].
\ee
$M,N$ run over $0,1,2,3,4$, while $\mu, \nu$ run over $0,1,2,3$.
Our conventions on the gauge group generators and the metric are
$\text{Tr}(T^{\cal A} T^{\cal B})=\frac 12 \delta^{{\cal A}{\cal B}}$
and $\eta_{MN}=\text{diag}(-1,1,1,1,1)$. The coupling $1/g^2_{5}$ has
been rescaled inside the covariant derivative, $D_{M}=
\partial_{M}+ig_{5} A_{M}$, where $A_{M}$ is a standard gauge
vector field and $F_{MN}$ its field strength.
The other fields are a real scalar $\Sigma$, an $SU(2)_{R}$ triplet of
real auxiliary fields $X^{a}$, $a=1,2,3$ and a symplectic Majorana
spinor $\gl_{i}$ with $i=1,2$ which form an $SU(2)_R$ doublet. The reality condition is
\be
\gl^i= \epsilon^{ij} C\bar{\gl}_{j}^{T} 
\label{real2}
\ee
where $\epsilon^{12}=1$ and $C$ is the 5d charge conjugation matrix
$C\gamma^M C^{-1}=(\gamma^M)^T$. An explicit realisation of the
Clifford algebra $\{\gamma^M,\gamma^N\}=-2\eta^{MN}$ is
\be
\gamma^M=\left(\,\left(\begin{array}{cc}0&\sigma^\mu_{\alpha \dot{\alpha}}\\ 
\bar{\sigma}^{\mu \dot{\alpha} \alpha }&0
\end{array}\right),
\left(\begin{array}{cc}-i&0\\ 0&i\end{array}\right)\,
\right)\,,~~\mbox{and}~~~
C=\left(\begin{array}{cc}
-\epsilon_{\alpha\beta} & 0\\ 
0 & \epsilon^{\dot\alpha \dot\beta}
\end{array}\right)\,,
\ee
where $\sigma^\mu_{\alpha \dot{\alpha}}=(1,\vec{\sigma})$ and
$\bar{\sigma}^{\mu \dot{\alpha}
\alpha}=(1,-\vec{\sigma})$. $\alpha,\dot{\alpha}$ are spinor indices
of $\text{SL}(2,C)$. This action is supersymmetric under the susy
transformations
\bea
\delta_{\epsilon}A^M &=& i\bar{\epsilon}_i\gamma^M \gl^{i}\\
\delta_{\epsilon}\Sigma &=& i\bar{\epsilon}_i \gl^{i}\\
\delta_{\epsilon} \gl^{i} &=& (\gamma^{MN} F_{MN}-\gamma^M D_{M}\Sigma)
\epsilon^i -i(X^a \sigma^a)^{i}_{~j} \epsilon^j\\
\delta_{\epsilon}X^a &=& \bar{\epsilon}_i(\sigma^a)^{i}_{~j}
\gamma^{M} D_{M}\gl^j -ig_{5} 
[\Sigma,\bar{\epsilon}_i(\sigma^a)^i_{~j}\gl^j]
\eea
with $\gamma^{MN}=\frac{1}{4}[\gamma^M,\gamma^N]$. The symplectic
Majorana spinor supersymmetry parameter is
$\bar{\epsilon}_i=\epsilon_{i}^{\dagger}\gamma^{0}$. To clarify
notation we temporarily display all labels, writing the Dirac spinor
in two component form ${\psi}^{i \, T}= (\psi_{\alpha}^{L i},
\bar{\psi}^{R \dot{\alpha} i})$ and $\bar{\psi}_i= (\psi^{R \alpha}_{
i}, \bar{\psi}^L_{ \dot{\alpha} i})$.  The bar on the two component
spinor denotes the complex conjugate representation of $SL(2,C)$.  
In particular, the reality condition~\eqref{real2} implies that
\be
\lambda^1 = \left(\begin{array}{c}
\lambda_{L \alpha}\\ 
\bar\lambda_{R}^{\dot\alpha}
\end{array}\right)~,~~~
\lambda^2 = \left(\begin{array}{c}
\lambda_{R \alpha}\\ 
-\bar\lambda_{L}^{\dot\alpha}
\end{array}\right)~,~~~
(\bar\lambda_1)^{T} = \left(\begin{array}{c}
\lambda_{R}^{\alpha}\\ 
\bar\lambda_{L \dot\alpha}
\end{array}\right)~,~~~
(\bar\lambda_2)^{T} = \left(\begin{array}{c}
-\lambda_{L}^{\alpha}\\ 
\bar\lambda_{R \dot\alpha}
\end{array}\right)~,
\ee
so the $SU(2)_{R}$ index on a two component spinor is a redundant
label.

Next, using an orbifold $S^{1}\!/\mathbb{Z}_{2}$ the boundaries will
preserve only half of the $\mathcal{N}=2$ symmetries. We choose to
preserve $\epsilon_{L}$ and set $\epsilon_{R}=0$. The conjugate
representations are constrained by the reality condition \refe{real2}.
The susy transformations are
\bea
\delta_{\epsilon_L}A^\mu &=& i\bar{\epsilon}_L\bar{\sigma}^\mu\lambda_L+i\epsilon_L\sigma^\mu
\bar{\lambda}_L\label{delam}\\
\delta_{\epsilon_L}A^5 &=& -\bar{\epsilon}_L\bar{\lambda}_R-\epsilon_L\lambda_R\\
\delta_{\epsilon_L}\Sigma &=& i\bar{\epsilon}_L\bar{\lambda}_R-i\epsilon_L\lambda_R\\
\delta_{\epsilon_L}\lambda_L &=& \sigma^{\mu \nu}F_{\mu \nu}\epsilon_L-iD_5\Sigma\epsilon_L
+iX^3\epsilon_L\\
\delta_{\epsilon_L}\lambda_R &=& i\sigma^\mu F_{5\mu}\bar{\epsilon}_L-\sigma^\mu D_\mu
\Sigma\bar{\epsilon}_L+i(X^1+iX^2)\epsilon_L\\
\delta_{\epsilon_L}(X^1+iX^2) &=& 2\bar{\epsilon}_L\bar{\sigma}^\mu D_\mu \lambda_R-2i
\bar{\epsilon}_LD_5\bar{\lambda}_L+ig_{5}[\Sigma,2\bar{\epsilon}_L\bar{\lambda}_L]\\
\delta_{\epsilon_L} X^3 &=& \bar{\epsilon}_L\bar{\sigma}^\mu D_\mu \lambda_L+i
\bar{\epsilon}_LD_5\bar{\lambda}_R-\epsilon_L\sigma^\mu D_\mu\bar{\lambda}_L
-i\epsilon_LD_5\lambda_R\nonumber\\
&&+ig_{5}[\Sigma,(\bar{\epsilon}_L\bar{\lambda}_R+\epsilon_L\lambda_R)]\,,\label{delx3}
\eea
where $\sigma^{\mu \nu}=\frac{1}{4}(\sigma^{\mu}
\bar{\sigma}^{\nu}-\sigma^{\nu}\bar{\sigma}^{\mu} )$. We have a parity
operator $P$ of full action $\mathbb{P}$ and define
$P\psi_{L}=+\psi_{L}$ $P\psi_{R}=-\psi_{R}$ for all fermionic fields
and susy parameters\footnote{The assignment
$P\partial_{5}=-\partial_{5}$ is also required.}. One can group the
susy variations under the parity assignment and it becomes clear that
the even parity susy variations are those of an off-shell $4d$ vector
multiplet $V(x_{5})$.  Similarly the susy variations of odd parity
form a chiral superfield $\Phi(x_{5})$. We may therefore write a $5d$
$\mathcal{N}=1$ vector multiplet as a $4d$ vector multiplet and a
chiral superfield:
\begin{alignat}{1}
V=&- \theta\sigma^{\mu}\bar{\theta}A_{\mu}+i\bar{\theta}^{2}\theta\gl-
i\theta^{2}\bar{\theta}\bar{\gl}+\frac{1}{2}\bar{\theta}^{2}\theta^{2}D\\
\Phi=& \frac{1}{\sqrt{2}}(\Sigma + i A_{5})+
\sqrt{2}\theta \chi + \theta^{2}F\,,
\label{fields2}
\end{alignat}
where the identifications between $5d$ and $4d$ fields are
\begin{equation}
D=(X^{3}-D_{5}\Sigma) \quad F=(X^{1}+iX^{2})\,,
\end{equation}
and we used $\lambda$ and $\chi$ to indicate $\lambda_{L}$ and
$-i\sqrt{2}\lambda_R$ respectively. The Non-Abelian bulk action in ${\cal N}=1$  4D formalism is
\begin{equation}
S^{SYM}_{5}= \int d^{5}x \left\{\frac{1}{2}\text{Tr} \left[ \int d^{2}\theta
 W^{\alpha}W_{\alpha}+
\int d^{2}\bar{\theta}  \bar{W}_{\dot{\alpha}}\bar{W}^{\dot{\alpha}}\right]  
+ \frac{1}{2 g_{5}^2} \int d^{4}\theta  
\text{Tr}\left[e^{-2g_{5}V}\nabla_5 e^{2g_{5}V}\right]^2\right\}\, .
\label{fullaction}
\end{equation}
 $\nabla_5$ is a ``covariant'' derivative with the respect to the
field $\Phi$~\cite{Hebecker:2001ke}:
\be \nabla_5 e^{2g_{5}V}=\partial_5
e^{2g_{5}V} - g_{5}\Phi^\dagger e^{2g_{5}V} - g_{5} e^{2g_{5} V} \Phi . 
\ee

Let us now focus on 5d hypermultiplets. The bulk supersymmetric action is 
\bea
S_{5D}^{H} &=& \int d^5 x[-(D_MH)^\dagger_i(D^MH^i)-
i\bar{\psi}\gamma^MD_M\psi+ F^{\dagger i}F_i-g_5
\bar{\psi}\Sigma\psi+g_5 H^\dagger_i(\sigma^aX^a)^i_jH^j
\nonumber\\
&& +g_5^2 H^\dagger_i\Sigma^2H^i+ig_5\sqrt{2}\bar{\psi}
\lambda^i\epsilon_{ij} H^j-i\sqrt{2}g_5H^{\dagger}_{i}
\epsilon^{ij}\bar{\gl}_{j}\psi\,].\label{hyperaction3}
\eea
$H_{i}$ are an $SU(2)_{R}$ doublet of scalars. $\psi$ is a Dirac
fermion and $F_{i}$ are a doublet of scalars. With our conventions,
the dimensions of ($H_{i},\psi,F_{i}$) are
($\frac{3}{2},2,\frac{5}{2}$). In general the hypermultiplet matter will be in a representation of the gauge group with Dynkin index defined by $d\delta^{ab}=\text{Tr}[T^a T^b]$. The action is supersymmetric under the susy transformations
\bea
\delta_\epsilon H^i &=& -\sqrt{2}\epsilon^{ij}\bar{\epsilon}_j\psi\label{delh}\\
\delta_\epsilon \psi &=& ig_5\sqrt{2}\gamma^MD_MH^i\epsilon_{ij}\epsilon^j-g_5\sqrt{2}\Sigma
H^i\epsilon_{ij}\epsilon^j+\sqrt{2}F_i\epsilon^i\\
\delta_\epsilon F_i &=& i\sqrt{2}\bar{\epsilon}_i\gamma^MD_M\psi+g_5\sqrt{2}\bar{\epsilon}_i
\Sigma\psi-2ig_5\bar{\epsilon}_i\lambda^j\epsilon_{jk}H^k\,.
\eea
To obtain the $\mathcal{N}=1$ sets due to the boundaries preserving only half the supersymmetry, we again choose to preserve $\epsilon_{L}$ and set $\epsilon_{R}=0$.  The susy variations are
\bea
\delta_{\epsilon_{L}} H^1 &=& \sqrt{2}\epsilon_{L}\psi_{L} \\
\delta_{\epsilon_{L}} H^2 &=& \sqrt{2}\bar{\epsilon}_{L}\bar{\psi}_{R} \\
\delta_{\epsilon_{L}} \psi_{L\alpha} &=& ig_5\sqrt{2}\sigma^{\mu}_{\alpha \dot{\beta}}D_\mu H^2\bar{\epsilon}^{L\dot{\beta}} +g_5\sqrt{2}D_{5}H^2 \epsilon^{L}_{\alpha}- g_5\sqrt{2}\Sigma
H^2\epsilon^{L}_{\alpha}+\sqrt{2}F_1\epsilon^{L}_{\alpha}\\
\delta_{\epsilon_{L}} \bar{\psi}^{R \dot{\alpha}} &=& 
ig_5\sqrt{2}\bar{\sigma}^{\mu\dot{\alpha}\beta}D_\mu H^2\epsilon^{L}_{\beta}
-g_5\sqrt{2}D_{5}H^1\bar{\epsilon}^{L \dot{\alpha}}
-g_5\sqrt{2}\Sigma
H^1\bar{\epsilon}^{L\dot{\alpha}}
-\sqrt{2}F_2\bar{\epsilon}^{L\dot{\alpha}}\\
\delta_{\epsilon_{L}} F_1 &=& i\sqrt{2}\bar{\epsilon}_{L\dot{\alpha}}\bar{\sigma}^{\mu \dot{\alpha}\beta}D_\mu \psi_{L \beta}
-\sqrt{2}\bar{\epsilon}^{L}_{\dot{\alpha}}D_{5}\bar{\psi}^{R \dot{\alpha}}+g_5\sqrt{2}\bar{\epsilon}^{L}_{\dot{\alpha}}\Sigma \bar{\psi}^{R \dot{\alpha}}-2ig_5\bar{\epsilon}^{L j}_{ \dot{\alpha}}\bar{\lambda}^{R\dot{\alpha}j}\epsilon_{jk}H^k\\
\delta_{\epsilon_{L}} F_2 &=& -i\sqrt{2} \epsilon^{L\alpha} 
\sigma^{\mu}_{\alpha\beta}D_\mu \bar{\psi}^{R \dot{\beta}}
-\sqrt{2}\epsilon^{L \alpha}D_{5}\psi^{L}_{\alpha}-g_5\sqrt{2}\epsilon^{L \alpha}\Sigma \psi^{L}_{\alpha} +2ig_5\epsilon^{L \alpha}\lambda^{L \alpha j}\epsilon_{jk}H^k\,.
\eea
In the 4d superfield formulation, we again use the parity of the $P\psi_{L}=+\psi_{L}$ and $P\psi_{R}=-\psi_{R}$ to group the susy transformations into a positive and negative parity chiral superfields, $PH=+H$ and $PH^c=-H^c$:
\bea
H &=& H^1+\sqrt{2}\theta\psi_L+\theta^2(F_1+D_5H_{2}-g_5\Sigma H_2)\\
H^c &=& H^\dagger_2+\sqrt{2}\theta\psi_R+\theta^2(-F^{\dagger}_{2}-D_5
H^\dagger_1- g_5 H^\dagger_1\Sigma)\,.
\eea
The gauge transformations are
$H\to e^{-\Lambda}H$ and $H^c\to H^ce^\Lambda$. The $\mathcal{N}=1$ action in 4d language is
\be
S_{5d}^H=\int d^5 x (\int d^4\theta [ H^\dagger e^{2g_5 V}H+H^c e^{-2g_5 V}H^{c\dagger}] + \int d^2 \theta  H^c\nabla_5 H+ \int d^2\bar{\theta}H^{c\dagger}\nabla_5 H^\dagger )
\,.
\ee

\section{Generalised messenger sector in 5D with an orbifold}\label{generalised}

This section extends the results of \cite{Marques:2009yu} to the case of bulk propagation in 5D with an orbifold.  We keep as close as possible to the notation of \cite{Marques:2009yu}.  We consider a messenger sector $\phi_{i},\tilde{\phi}_{i}$ coupled to a SUSY breaking spurion $X$:
\be W = {\cal M}(X)_{ij}\ \phi_i \ti \phi_j =  (m + X \lambda)_{ij}\
\phi_i \ti \phi_j \label{superpotential2}\ee
$m$ and $\gl$ are generic matrices.  The messengers are in a representation of the gauge group with a Dynkin index $d$, defined by $d\delta^{ab}=\text{Tr}[T^a T^b]$.
The fundamental messengers on the SUSY breaking brane will couple to the bulk vector superfield as
\be \delta {\cal L} = \int d^2\theta d^2\bar\theta \left(\phi^\dag_i
e^{2 g V^a T^a} \phi_i + \ti\phi^\dag_i e^{-2 g V^a T^a}
\ti\phi_i\right) + \left(\int d^2\theta\ W +
c.c.\right) \label{hiddensector} \ee
We can extract the multiplet of currents from the kinetic terms in the above Lagrangian.  We find 
\be {\cal J}^a = J^a + i \theta j^a - i \bar \theta  \bar j^a -
\theta\sigma^\mu\bar \theta j_\mu^a +
\frac{1}{2}\theta\theta\bar\theta\bar \sigma^\mu\partial_\mu j^a -
\frac{1}{2}\bar\theta\bar\theta\theta \sigma^\mu\partial_\mu \bar
j^a - \frac{1}{4}\theta\theta\bar\theta\bar\theta \Box J^a\ee
where
\bea
J^a &=& \phi_i^\dag T^a \phi_i - \ti \phi^\dag_i T^a \ti \phi_i \nn\\
j^a &=& - i \sqrt{2} \left(\phi_i^\dag T^a \psi_i - \ti \phi^\dag_i T^a \ti \psi_i\right) \nn\\
\bar j^a &=& i \sqrt{2} \left(\bar \psi_i T^a \phi_i - \bar{\ti \psi}_i T^a \ti \phi_i\right) \nn\\
j_\mu^a &=& \left(\psi_i \sigma_\mu T^a \bar \psi_i - \ti \psi_i
\sigma_\mu T^a \bar{\ti\psi}_i\right) - i \left( \phi_i^\dag T^a
\partial_\mu \phi_i - \partial_\mu\phi^\dag_i T^a \phi_i -
\ti \phi_i^\dag T^a\partial_\mu \ti \phi_i + \partial_\mu\ti
\phi^\dag_i T^a \ti \phi_i\right)   
\label{supercurrent}\eea

(repeated indices are summed)
\bea
\ti C_0 &=&\sum_{k,n} 2 d_{kn} B_{kn} \int \frac{d^4q}{(2\pi)^4} \frac{1}{(q^2 + (m_k^+)^2)((p + q)^2 + (m_n^-)^2)} \\
\ti C_{1/2} &=& -\sum_{k,n} \frac{2 d_{kn}}{p^2} \sum_\pm  A^\pm_{kn} \int
\frac{d^4q}{(2\pi)^4} \frac{p\cdot q}{((p
+ q)^2 + (m_k^\pm)^2)(q^2 + (m_n^0)^2)} \\
\ti C_{1} &=& - \sum_{k,n}\frac{2 d_{kn}}{3 p^2} \int \frac{d^4q}{(2\pi)^4}\
\delta_{kn} \left[\sum_\pm \left(\frac{(p + q)\cdot (p + 2q)}{(q^2 +
(m_k^\pm)^2)((p + q)^2 + (m_k^\pm)^2)}- \frac{4}{q^2 +
(m_k^\pm)^2}\right) \right.\nn\\
&& \left. \ \ \ \ \ \ \ \ \ \ \ \ \ \ \ \ \ \ \ \ \ \ \ \ \ \ \ +
\frac{4 q\cdot(p + q) + 8 (m_k^0)^2}{(q^2 + (m_k^0)^2)((p
+ q)^2 + (m_k^0)^2)} \right] \\
M \ti B_{1/2} &=& \sum_{k,n}2 d_{kn} \sum_\pm \mp A^\pm_{kn} \int
\frac{d^4q}{(2\pi)^4} \frac{m_n^0}{(q^2 + (m_k^\pm)^2)((p +
q)^2 + (m_n^0)^2)} \eea
The $\ti C_a$ may be written as
\bea
 \ti C_0
\ &=& \ \sum_{k,n} 2 d_{kn} B^+_{kn} G_1(m_k^+, m_n^-) \\
-4 \ti C_{1/2}\ &=& -\sum_{k,n}\ 4 d_{kn} \sum_\pm A^\pm_{kn}\left[ (G_0(m_k^\pm) -
G_0(m_n^0))+ G_1 (m_k^\pm, m_n^0) \right.\label{cs}\\&& \ \ \ \ \ \ \ \ \ \
\ \ \ \ \ \ \left.  + ((m_k^\pm)^2 - (m_n^0)^2)\frac{1}{p^2}G_1(m_k^\pm, m_n^0)\right]\nn\\
  3 \ti C_1 \ &=&\sum_{k,n}
\  d_{kn} \delta_{kn} \sum_\pm \left[4 (G_0(m_k^\pm) - G_0 (m_k^0)) + G_1(m_k^\pm, m_k^\pm) \right. \\
&&\ \ \ \ \ \ \ \ \ \ \ \ \  \left.+ 2 G_1(m_k^0, m_k^0) +\ 4 (m_k^\pm)^2 \frac{1}{p^2}G_1 (m_k^\pm, m_k^\pm) -
4 (m_k^0)^2 \frac{1}{p^2}G_1(m_k^0, m_k^0)\right] \nn
 \eea
All the terms proportional to $G_0$ in (\ref{cs}) vanish due to the messenger supertrace formula.  The G functions are
\bea G_0(m) &=& \int \frac{d^4 p}{(2\pi)^4} \frac{1}{p^2 + m^2} \label{G0}\\
G_1(m_1, m_2) &=& \! \! \int\!
\frac{d^4 q}{(2\pi)^4}  \frac{1}{(q^2 + m_1^2)((p + q)^2 +
m_2^2)}\label{G1}\label{G3}
\eea
We then define the function $b(k^2,m_1^2,m_2^2)$ by 
\begin{equation}
 G_1(m_1, m_2)=   \int \frac{d^4 q}{(2\pi)^4}\frac{1}{ q^2 + m_1^2}\frac{1}{ (p + q)^2 + m_2^2}
 = \frac{1}{ (4\pi)^2}\big\{ \frac{1}{ \epsilon} - \gamma - b(p^2,m_1^2,m_2^2) + 
        {\cal O}(\epsilon)\big\}
\label{expandsc}\end{equation}
for $d = 4-\epsilon$.  We can write $b$ more
explicitly as
\begin{eqnarray}
b(p^2,m_1^2,m_2^2) &=& \int^1_0 dx \log\left(x(1-x)p^2 + xm_1^2 + (1-x)
  m_2^2\right)\nonumber \\ 
 & = & A \log\left[\frac{(A + B_1)(A+ B_2)}{ (A-B_1)(A-B_2)}\right]
        + B_2 \log m_1^2 + B_1\log m_2^2 - 2  \ ,
\label{valofb}\end{eqnarray}
where 
\begin{equation}
 A = \left[\frac{p^4 + 2p^2(m_1^2+m_2^2) + (m_1^2-m_2^2)^2}{ 4p^4}\right]^{1/2}
\label{Avalue}\end{equation}
and
\begin{equation}
 B_1 = \frac{p^2 + m_1^2 - m_2^2}{ 2p^2}\ , \qquad 
 B_2 = \frac{p^2 + m_2^2 - m_1^2}{ 2p^2}\ . 
\label{Bvalue}\end{equation}
As the divergent terms will cancel we can ignore the Euler $\gamma$ term and $\epsilon$ and focus on the $b$ functions.  We can rewrite the integral as
\be
G_1(m_1, m_2) = -\frac{1}{(4\pi)^2}b(p^{2},m_{1}^{2},m_{2}^{2})+ g(\epsilon, \gamma) \label{G32}
\ee
Where we can safely ignore the functions $g(\epsilon, \gamma)$ as they will cancel.  Putting this together we can write
\bea [3\ti C_{1}-4\ti C_{1/2}+ \ti C_{0} ] &=& -[\sum_{k,n}  2 d_{kn} B^{+}_{kn} b(p^{2},m_{k}^{+ 2},m_{n}^{- 2})]\nn \\
&+&\! \!  \sum_{k,n} 4 d_{kn} \sum_\pm A^\pm_{kn}[b(p^{2},m_{k}^{\pm 2},m_{n}^{0 2})+\frac{1}{p^{2}}((m^{\pm}_{k})^{2}-(m^{0}_{n})^{2}) b(p^{2},m_{k}^{\pm 2},m_{n}^{0 2})]\nn \\
&-&\sum_{k,n} d_{kn} \delta_{kn} \sum_\pm[b(p^{2},m_{k}^{\pm 2},m_{k}^{\pm 2})+2b(p^{2},m_{k}^{0 2},m_{k}^{0  2})  \\
&&\ \ \ \  +4\frac{1}{p^{2}}(m^{\pm}_{k})^{2}b(p^{2},m_{k}^{\pm 2},m_{k}^{\pm  2})-4\frac{1}{p^{2}}(m^{0}_{k})^{2}b(p^{2},m_{k}^{0 2},m_{k}^{0  2})]\nn \label{useful}
\eea
We may construct a dictionary of 
\be 
[3\tilde{C}_1^{(r)}(p^2/M^2)-4\tilde{C}_{1/2}^{(r)}(p^2/M^2)+\tilde{C}_0^{(r)}(p^2/M^2)]= \frac{1}{(4\pi)^2}\Xi
\ee expressions depending on the hidden sector,  such that they may be used both in 4d and various 5d models.  The factors of two and $\pi$ on the right hand side can simply be taken out to convert $g^2 \rightarrow \alpha$ when going from \refe{mainresult} to \refe{alpha}
We now need to find limits of the b function, which we outline in the next subsection.  

The Majorana gaugino mass matrix couples \emph{every} Kaluza-Klein mode to \emph{every} other mode with the same coefficient.  Each entry can be determined by use of the $b$ function and is given as
\be M^{\ti g} = g^2 M \tilde{B}_{1/2}(0) =\frac{\alpha_r}{4 \pi}
\Lambda_{G} \ , \ \ \ \ \Lambda_{G} = 2 \sum_{k,n = 1}^N \sum_\pm \pm\ d_{kn}\ A_{kn}^\pm\ m^0_n 
 \frac{(m_k^\pm)^2 \log ((m_k^\pm)^2/(m_n^0)^2)}{(m_k^\pm)^2 - (m_n^0)^2}. \label{45}\ee
$k,n$ are messenger indices running from $1$ to $N$, the number of messengers, while $d_{kn}$ is nonzero and equal to $d_{k}$ or $d_{n}$ only if $\phi_{n}$ and $\tilde{\phi}_{k}$ are in the same representation. To find the mass eigenstates, one must include the Dirac masses resulting from the Kaluza-Klein tower.

\subsection{Expansions of b}
We may expand \refe{valofb}  under different limits to get practical expressions which one may then substitute into \refe{useful}.
In the large $\ell$ limit the integral over $p$ in~\refe{mainresult} receives a sizeable contribution only from the region of small momenta ($p<1/\ell$), while the remaining part of the region of integration is exponentially suppressed. So we can expand the function $b$ in~\refe{valofb}, which is the building block for the full integrand~\eqref{useful}, for small momenta. In the regime $1/\ell^2\ll F,M^2$, this expansion gives
\be 
b(p^2,m^2_1,m^2_2)\approx -1 +\frac{m^2_1\log m^2_1-m^2_2 \log m^2_2}{m^2_1-m^2_2}+
\frac{p^2}{2(m^2_1-m^2_2)^{3}}[m^4_{1}-m^4_{2}-2m^2_{1}m^2_{2}\log \frac{m^2_{1}}{m^2_{2}}]\label{nopoles}~.
\ee
Even if it is not immediately obvious, there are no poles in this equation \refe{nopoles} when $F\rightarrow 0$ (i.e. $m_{1}\rightarrow m_{2}$).  Using this limit we can obtain the expression for when both masses are equal:
\be\label{lellem}
 b(p^2,m^2,m^2)\approx \log m^2 +\frac{p^2}{6m^2}+O(\frac{p^4}{m^4}),
\ee
which shows that there are no discontinuities when $F$ changes from $F\gg 1/\ell^2$ to $F\leq 1/\ell^2$. \label{sec:cterms}
\subsection{Minimal GMSB}
When the superpotential is of the form 
\be
W= X\Phi_{i}\bar{\Phi}_{i} 
\ee
where $X=M+\theta^{2}F$, then the model only has three masses: ($M^{2}, M^{2}_{+}=M^{2}(1+x),M^{2}_{-}=M^{2}(1-x)$).  The limit of small $\ell$, with flat space, the 5d model will return to the 4d result:
 \be 
 \Xi \approx -d\frac{4M^{4}}{p^{4}}[x^{2}\log(\frac{p^2}{M^2})+(x^2+3x+2)\log(1+x)-x^2+(x\rightarrow-x)]+O(p^{-6}).
 \ee
The intermediate limit $F\leq 1/\ell^2 \ll M^2$ may be found by expanding the large $\ell$ limit and taking the first order in $x=F/M^2$we find
 \be \Xi \approx- d  
\frac{2 F^2}{3 M^4}
 \ee 
In the large $\ell$ limit where $1/\ell^2 \ll F,M^2$ we find

\be 
\Xi \approx-d[\frac{4+x-2x^2}{x^2}\log(1+x)+1 +(x\rightarrow -x)]+O(p^{4}),
 \ee
In agreement with \cite{Mirabelli:1997aj}.  In this limit we may also make the identification
\be
\frac{3}{2} \Xi=-d x^2 h(x)
\ee
Where $h(x)$ is defined in section \ref{sfermionmasses}

 \subsection{Generalised Messenger Sector}
 When the superpotential of the hidden sector is 
 \be W = {\cal M}(X)_{ij}\ \phi_i \ti \phi_j =  (m + X \lambda)_{ij}\
\phi_i \ti \phi_j \ee
We have three cases. The limit of small $\ell$, with flat space, the 5d model will return to the 4d result\cite{Marques:2009yu}.
In the large $\ell$ limit such that $1/\ell^2 \ll F,M^2$, we can use \refe{nopoles} and \refe{lellem} and $\sum_{n}A_{nn}=\sum_{n}A_{kn}=1$ and $B^{+}_{kn}=B^{-}_{nk}$ to obtain \refe{LambdaS2}.

\providecommand{\href}[2]{#2}\begingroup\raggedright\endgroup

\end{document}